\documentclass[twocolumn,aps,prb,showpacs,preprintnumbers,amsmath,amssymb, superscriptaddress]{revtex4-1}
\usepackage{bm}
\usepackage{graphicx}
\usepackage{color,xspace}
\usepackage{ulem}
\usepackage{float}
\usepackage{textcomp}

\newenvironment{addendum}{%
   \setlength{\parindent}{0in}%
   \small%
   \begin{list}{Acknowledgements}{%
       \setlength{\leftmargin}{0in}%
       \setlength{\listparindent}{0in}%
       \setlength{\labelsep}{0em}%
       \setlength{\labelwidth}{0in}%
       \setlength{\itemsep}{12pt}%
       }
   }
   {\end{list}\normalsize}

\begin{document}

\title{Observation and control of the weak topological insulator state in ZrTe$_5$ }

\author{Peng~Zhang}
\email{zhangpeng@issp.u-tokyo.ac.jp}
\affiliation{Institute for Solid State Physics, University of Tokyo, Kashiwa, Chiba 277-8581, Japan}

\author{Ryo~Noguchi}
\affiliation{Institute for Solid State Physics, University of Tokyo, Kashiwa, Chiba 277-8581, Japan}

\author{Kenta~Kuroda}
\affiliation{Institute for Solid State Physics, University of Tokyo, Kashiwa, Chiba 277-8581, Japan}

\author{Chun~Lin}
\affiliation{Institute for Solid State Physics, University of Tokyo, Kashiwa, Chiba 277-8581, Japan}

\author{Kaishu~Kawaguchi}
\affiliation{Institute for Solid State Physics, University of Tokyo, Kashiwa, Chiba 277-8581, Japan}

\author{Koichiro~Yaji}
\affiliation{Research Center for Advanced Measurement and Characterization, National Institute for Materials Science, Tsukuba, Ibaraki 305-0003, Japan}

\author{Ayumi~Harasawa}
\affiliation{Institute for Solid State Physics, University of Tokyo, Kashiwa, Chiba 277-8581, Japan}

\author{Mikk~Lippmaa}
\affiliation{Institute for Solid State Physics, University of Tokyo, Kashiwa, Chiba 277-8581, Japan}

\author{Simin~Nie}
\affiliation{Department of Materials Science and Engineering, Stanford University, Stanford, California 94305, USA}

\author{Hongming~Weng}
\affiliation{Beijing National Laboratory for Condensed Matter Physics and Institute of Physics, Chinese Academy of Sciences, Beijing 100190, China}

\author{V.~Kandyba}
\affiliation{Elettra - Sincrotrone Trieste, Basovizza, Italy}

\author{A.~Giampietri}
\affiliation{Elettra - Sincrotrone Trieste, Basovizza, Italy}

\author{A.~Barinov}
\affiliation{Elettra - Sincrotrone Trieste, Basovizza, Italy}

\author{Qiang~Li}
\affiliation{Condensed Matter Physics and Materials Science Department, Brookhaven National Laboratory, Upton, NY, USA}

\author{G.D.~Gu}
\affiliation{Condensed Matter Physics and Materials Science Department, Brookhaven National Laboratory, Upton, NY, USA}

\author{Shik~Shin}
\affiliation{Office of University Professor, University of Tokyo, Kashiwa, Chiba 277-8581, Japan}
\affiliation{Institute for Solid State Physics, University of Tokyo, Kashiwa, Chiba 277-8581, Japan}

\author{Takeshi~Kondo}
\email{kondo1215@issp.u-tokyo.ac.jp}
\affiliation{Institute for Solid State Physics, University of Tokyo, Kashiwa, Chiba 277-8581, Japan}

\date{\today}

\maketitle

\textbf{A quantum spin Hall (QSH) insulator hosts topological states at the one-dimensional (1D) edge, along which  backscattering by nonmagnetic impurities is strictly prohibited and dissipationless current flows \cite{KanePRL2005, ZhangScience2006, MolenkampScience2007,LiScience2014, JHScience2018}. Its 3D analogue,  a ``weak'' topological insulator (WTI), possesses similar quasi-1D topological states confined at ``side'' surfaces of crystals. The enhanced confinement could provide a route for dissipationless current and better advantages for applications relative to the widely studied ``strong'' topological insulators (STIs).  However, the topological side surface is usually not cleavable and is thus hard to observe by angle-resolved photoemission spectroscopy (ARPES), which has hindered the revealing of the electronic properties of WTIs \cite{FelserPRL2012, BrinkNM2013, WengPRX2014, YanPRB2014, ZhangPRL2016, PlucinskiNC2017, KondoNature2019}. Here, we visualize the topological  surface states of the WTI candidate ZrTe$_5$ for the first time by spin and angle-resolved photoemission spectroscopy: a quasi-1D band with spin-momentum locking was revealed on the side surface. We further demonstrate that the bulk band gap in ZrTe$_5$ is controlled by strain to the crystal, realizing a more stabilized WTI state or an ideal Dirac semimetal (DS) state depending on the direction of the external strain. The highly directional spin-current and the tunable band gap we found in ZrTe$_5$ will provide an excellent platform for applications.
}

Two of the most prominent examples of topological insulators in three dimensions are the STIs and WTIs \cite{FuPRB2007, FuPRB2007_2, HasanRMP2010, QiRMP2011}. Among them, STIs have been widely studied in the past decades both in theories and experiments. They host 2D spin-momentum locked Dirac cones on all surfaces, in which the perfect backscattering is prohibited, while general scattering off 180\textdegree~still exists\cite{XuePRL2009, KomoriPRL2014}.
On the other hand, WTIs host surface states only on particular side surfaces. They were thought to be weak (or not robust) since two adjacent layers in even-layer WTIs may couple with each other, leading to a topologically trivial phase. However, it was later found that the surface states of WTIs are actually robust owing to the delocalization of surface electrons \cite{RingelPRB2012}. 
The weak interlayer coupling in WTIs generally yields a topological surface state with quasi-1D dispersion; this could prohibit even general scattering off 180\textdegree~to establish dissipationless spin current, which cannot be realized in STIs.
Even with such advantages, theoretical proposals of WTIs are rare, and experimental investigations of the topological side surfaces are very challenging; layered WTI materials are inherently cleavable only on the top surface, and thus there is difficulty in preparing a large and uniform side surface for observation \cite{FelserPRL2012, BrinkNM2013, WengPRX2014, YanPRB2014, ZhangPRL2016, KondoNature2019}. One exception is Bi$_4$I$_4$, where top and side planes are both naturally cleavable \cite{KondoNature2019}. However, such crystal structures may pose obstacles in preparing a single side surface with the topological feature for applications.

\begin{figure*}[!thbp]
\begin{center}
\includegraphics[width=0.75\textwidth]{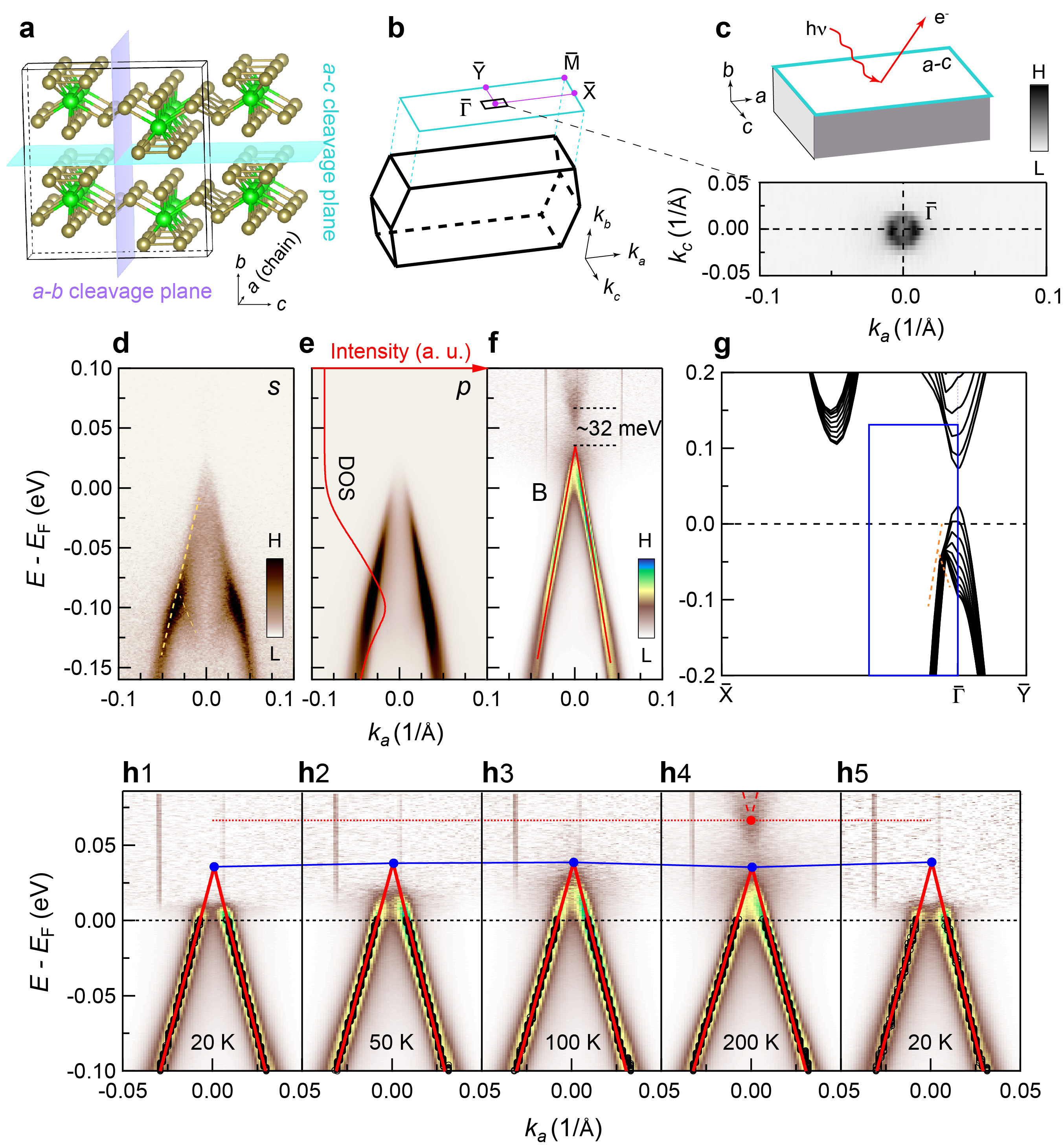}
\end{center}
 \caption{\label{topband} \textbf{Band structure of the top \textit{a-c} surface. } (a) Crystal structure of ZrTe$_5$ and its cleavage planes. (b) The bulk Brillouin zone (BZ) and its projected surface BZ on the top \textit{a-c} surface. (c) Fermi surface measured from the \textit{a-c} surface, of which the range is indicated by the black box in (b). (d - e) Band structure measured from the \textit{a-c} surface with $s$ and $p$-polarized light at 200 K, respectively. Light-yellow dashed lines in (d) highlight the band splitting around -0.1 eV. The red line in (e) shows the density of states (DOS).  (f) Same as (e), but normalized with its DOS for a clear presentation of the conduction band above Fermi level ($E_\mathrm{F}$), which will not affect the MDC peak positions. Details of the normalization can be found in Supplementary Information Part II and Fig. S2. (g) Band structure from first-principles slab calculations on the \textit{a-c} surface. The blue box roughly indicates the momentum range in (d - f).Orange dashed lines highlight the splitting similar to the one in (d). (h) Temperature dependence of the band structure taken with $p$-polarized light. The black markers are extracted from MDC peaks, and the red solid lines are the fitting results of the black markers. The blue line with solid markers indicates the top of the valence band at different temperatures. The red dashed line marks the conductance band bottom at 200 K. Details of the band fitting can be found in Methods. }
\end{figure*}

The material which has been best studied from the early stage of exploring a WTI among scarce candidates is ZrTe$_5$, which consists of a van der Waals layered structure; notably, this compound exhibits very high mobility \cite{ZhangNature2019}, thus it has been regarded as a promising platform for devices.  Nonetheless, the bulk topology of ZrTe$_5$ 
has not been experimentally identified to date because the observation of band structure on the side surface 
has not been successful so far. All the previous surface-sensitive studies were carried out on the top surface \cite{PanPRX2016,CrepaldiPRL2016, ZhouNC2017, ShenPRB2017}, which only confirms the lack of surface states; whether or not this material is topological, therefore, has not been determined beyond speculation.  
Another difficulty for this study is that ZrTe$_5$ is in proximity to multiple topological phases, whereas this feature could bring an attractive functionality of controlling bulk topology by fine-tuning a physical parameter \cite{WengPRX2014}.

\begin{figure*}[!thb]
\begin{center}
\end{center}
\includegraphics[width=.85\textwidth]{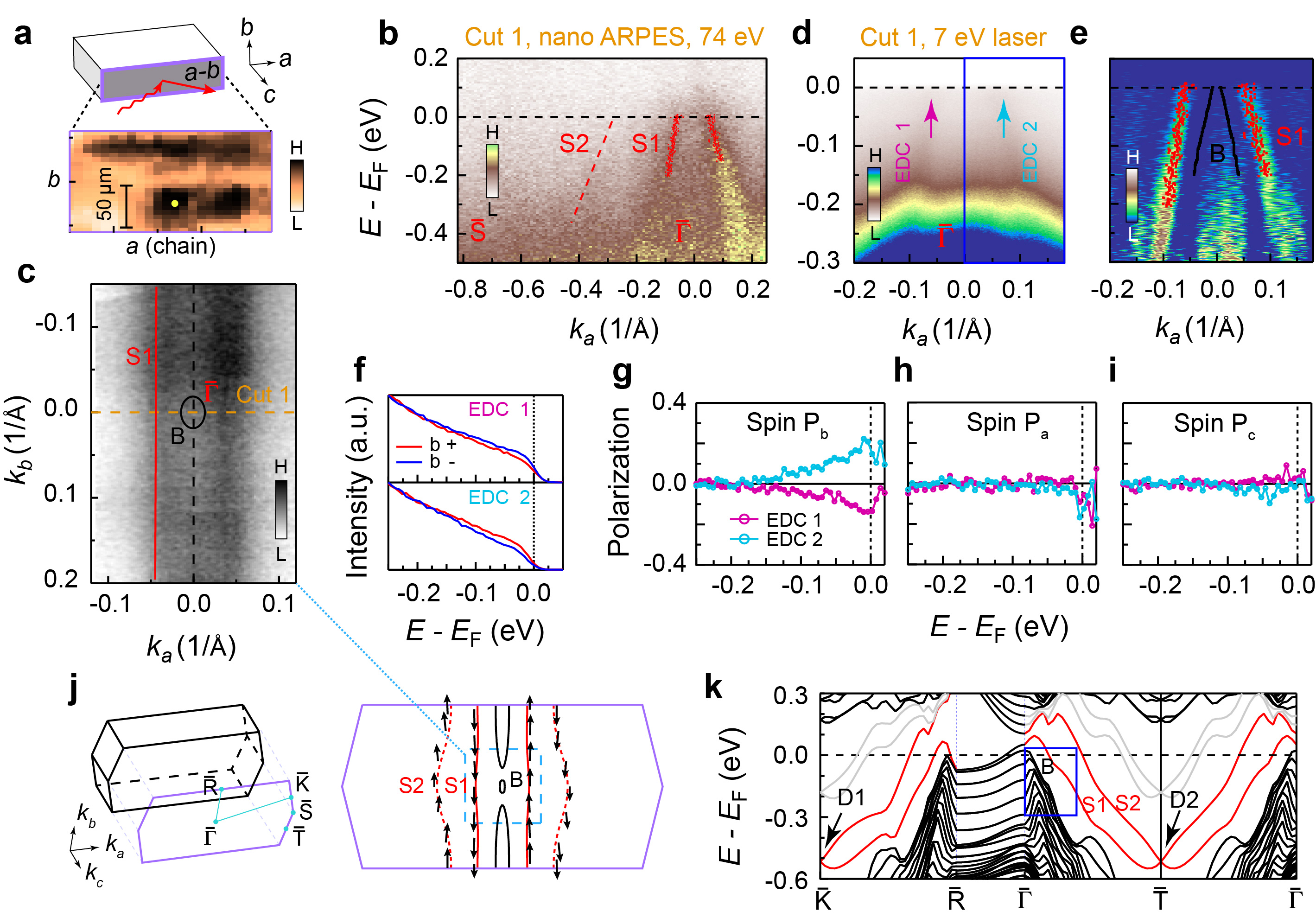}
\caption{\label{sideband} \textbf{Surface band of the side \textit{a-b} surface.} (a) Real space photoemission intensity mapping from a synchrotron-based nano-ARPES. The width of the cleaved area with high intensities of photoelectrons is about 50 $\mu$m. (b) Band structure from nano-ARPES with 74-eV photons at the position indicated by yellow solid dot in (a). The red dots are duplicates of the ones in (e), and the red dashed line roughly indicates the band S2 in (k).  (c) Fermi surface measured from the \textit{a-b} surface with laser ARPES. (d) Band structure of Cut 1 in (c). The blue box corresponds to the one in (k). (e) MDC curvature\cite{ZhangRSI2011} plot of (d). The red dots are extracted from the MDC peaks in (d), and the black dots (band B) are duplicates of the black markers in Fig. 1h4. (f) Spin-resolved ($b$ direction) EDCs at EDC 1 and EDC 2  indicated in (d), respectively. (g - i) Spin polarization curves along $a$, $b$ and $c$ directions at EDC 1 and EDC 2, respectively. More details on the spin data analysis can be found in Supplementary Information Part IV and Fig. S4. (j) Projected surface BZ and Fermi surface from first-principles slab calculations on the \textit{a-b} surface. The black arrows indicate the in-plane spin direction. (k) Band structure from first-principles slab calculations on the \textit{a-b} surface. The red lines indicate the surface bands, while the black lines represent the bulk bands. Due to the presence of the broken bonds at the surface, two trivial bands colored in grey appear below $E_\mathrm{F}$ in first-principles slab calculations. They are not intrinsic, and do not appear in the calculations based on Wannier functions \cite{WengPRX2014, ZhouSR2017}, of which the Fermi surfaces are not shown in (j).  The blue box corresponds to the one in (d).   }
\end{figure*}

\begin{figure*}[!htb]
\begin{center}
\includegraphics[width=.8\textwidth]{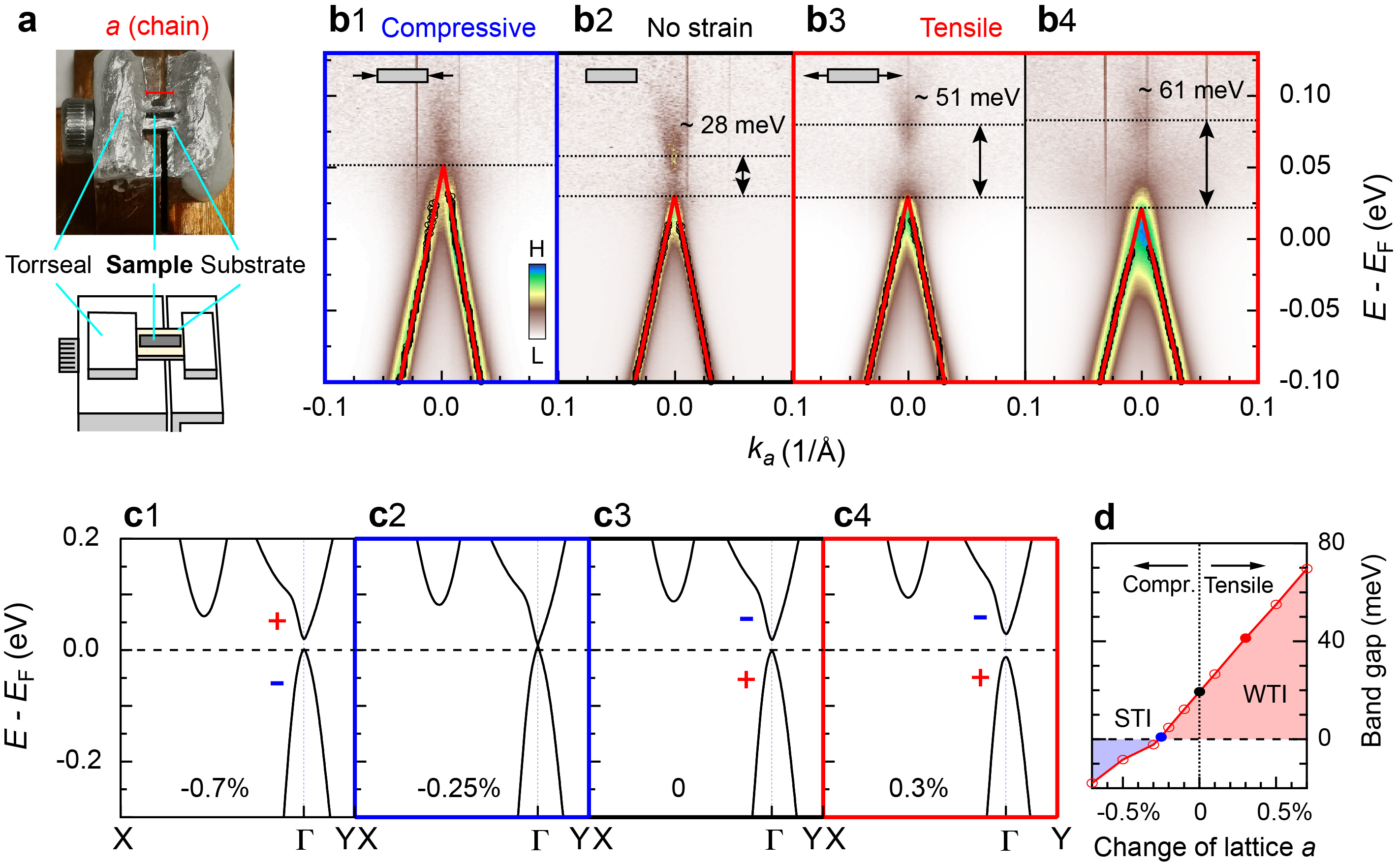}
\end{center}
 \caption{\label{pressure} \textbf{Control of the topological phase by external strain.} (a) The strain device. A screw is used to compress or stretch the substrate (and thus the sample attached to it) along the chain direction of the samples. The red line indicates the chain direction. (b) Bulk band gap change with compressive and tensile strain. With compressive strain (b1), the gap is (nearly) closed, reaching a Dirac semimetal state. With tensile strain (b3-b4), the band gap becomes larger, stabilizing the WTI state. The data are taken with $p$-polarized photons and normalized by their density of states (DOS). The black markers are extracted from the MDC peaks, and the red solid lines are the fitting results of the black markers, same as the ones in Fig. 1h. (c) Calculations on the band structure with different lattice constant $a$. + and -  signs indicate the inversion symmetry of the two bands.  In (b) and (c), The blue (red) frames correspond to compressive (tensile) strain. (d) Calculated phase diagram with different lattice constant (strain). Blue, black and red solid markers roughly indicate the experimental values in (b). }
\end{figure*}

ZrTe$_5$ has a quasi-1D crystal structure. The lattice constant $a$ along the chain direction is $\sim$ 4.0 \AA. Except for the chain direction, the lattice constants for the other two directions are both large, as shown in Fig. 1a: The layer distance along the $b$ direction is $\sim$ 7.3 \AA, and it is $\sim$ 6.9 \AA~along the $c$ direction \cite{FjellvagSSC1986}. As a result of the large layer distances, in principle, it should be possible to cleave both surfaces. Along the $b$ direction, the layers are stacked by van der Waals forces, and a clean surface can be easily obtained by cleavage. Instead, Te-Te bonding exists between adjacent layers along the $c$ direction, causing much more difficulty in cleaving the \textit{a-b} surface. All the spectroscopic studies reported so far have been carried out on the \textit{a-c} surface, which is easy to cleave \cite{VallaNP2016, PanPRX2016, LiPRL2016,CrepaldiPRL2016, ZhouNC2017, ShenPRB2017}. 

We observed a tiny hole-type Fermi surface on the \textit{a-c} surface, as shown in Fig. 1c. The band along $k_a$ (Fig. 1d-e) shows a cone-shape dispersion, similar to that in the previous reports except for the difference on Fermi level ($E_\mathrm{F}$) \cite{VallaNP2016, PanPRX2016, ZhouNC2017}. We notice a splitting of the band in Fig. 1d at $\sim$ -0.1 eV, with one branch going up and the other going down, as indicated by the dashed yellows lines in Fig. 1d. Such a feature most likely originates from a $k_z$ average effect, and it is indeed reproduced by the first-principles slab calculations plotted in Fig. 1g along $\overline{\Gamma}\overline{X}$, where the series of bands come from different $k_z$ ($k_b$); this further indicates that the slab calculations are reliable to explain the ARPES data. The measurements with $p$-polarized light (Fig. 1e) reveal only the cone-shaped branch going up, which is thus suitable for the study of the bulk band gap. 

To have a better visualization of the band structure over the whole energy range, the data in Fig. 1e is normalized by its density of states (DOS, integration of EDCs over $k_a$), as plotted in Fig. 1f (More details can be found in Supplementary Information Part II and Fig. S2). A small band gap with the size of about 32 meV is observed in the data at 200 K.  Our experiments over a temperature cycle of 20K-200K-20K (Fig. 1h) have confirmed almost no change in the valence band at different temperatures; this result clearly differs from the previous reports demonstrating a large energy shift in the band with temperature \cite{PanPRX2016, ZhouNC2017, ShenPRB2017, XuPRL2018}.  On the other hand, we have confirmed a band shift in an ARPES system with a worse vacuum level ($\sim 7 \times 10^{-11}$ Torr $v.s.$ $\sim 1 \times 10^{-11}$ Torr for Fig. 1h); the band shift actually occurs even for the sample kept at the same temperature  (See Supplementary Information Part I and Fig. S1 for more details). 
These results lead us to conclude that ZrTe$_5$ should be in a stable phase, which is not a strong topological insulator since surface states are absent in the data, but either a normal insulator or a WTI.  However, with the data merely from the top surface, it is impossible to distinguish the two phases experimentally.

To identify the topological phase of ZrTe$_5$, the observation of the side \textit{a-b} surface is essential.
We successfully cleaved the side surface with a top post and silver epoxy. The problem, however, is that the cleaved areas are very small and inhomogeneous. 
With a synchrotron-based nano-ARPES (spot size $<$ 1$\mu$m), we took real space maps of photoemission intensity on several samples, and found that the typical dimension of cleaved areas is around or smaller than 50 $\mu$m (Fig. 2a). Several positions of one cleaved area are measured and they all give the same band structure, as shown in Supplementary Information Part III and Fig. S3, indicating that the cleaved area with high photoelectron intensities is homogeneous. The selective measurements only of such a small surface area are beyond the capacity of most ARPES systems. Nevertheless, our laser-based micro-ARPES (spot size 50 $\mu$m) resolved this difficulty and clearly revealed the band structure on the side \textit{a-b} surface. 
A quasi-1D Fermi surface (S1) is exhibited in Fig. 2c by plotting ARPES intensities about the Fermi level. 
The band S1 shows a hole-like dispersion along $k_a$ (Fig. 2d-e), which has a Fermi vector $k_F$ much larger than that of the band B observed near $\overline{\Gamma}$ on the \textit{a-c} surface (Fig. 1f). The band S1 is, thus, distinct from the bulk band B located at the center of the BZ ($\overline{\Gamma}$). The bulk band B cannot be identified in ARPES data on the side \textit{a-b} surface likely due to a much weaker intensity than that of the surface states.  
The synchrotron-based nano-ARPES measurements for Cut1 with a different photon energy (Fig. 2b) show similar result to the data obtained by laser-ARPES: Only the band S1 was clearly observed, and its dispersion is almost identical to that of the laser-ARPES data. This coincidence supports that the band S1 has a surface origin. 

The spin polarization is expected for the topological surface state. To confirm this, we have used spin-resolved ARPES. 
In Fig. 2f, we plot EDCs (EDC 1 and EDC 2; see Fig. 2d) which are spin-resolved in the $b$-direction; 
a clear spin polarization is detected (Fig. 2g).  In contrast, the spin polarization is almost zero along the $a$ and $c$ directions (Fig. 2h-i).  Therefore, the band S1 is spin-polarized along the $b$-direction under the spin-momentum locking, forming a typical spin-texture of WTIs with quasi-1D dispersion (see Fig. 2j). 

We also carried out the first-principles slab calculations on the side \textit{a-b} surface to fully understand the ARPES data. The calculated Fermi surface (Fig. 2j) displays four sheets: two open sheets (red color) and two closed sheets (black color). By comparing these calculations with our ARPES data (Fig. 2c), we ascribe the inner open sheet to the band S1, and the closed sheets to the bulk band B. The outer open sheet (S2) was not observed with either laser-based ARPES or synchrotron-based ARPES most likely due to a small photoemission cross section. 
In the slab calculations, the bands S1 and S2 are localized at the surface and have spin polarization indicated by the black arrows in Fig. 2j; the spin orientation of the band S1 is consistent with our data of spin-resolved ARPES (Fig. 2f-i).
The slab calculations (Fig. 2k) clearly illustrate the topological nature of the bands S1 and S2, which connect the valence and conduction bands and cross with each other at two BZ boundary points, $\overline{K}$ and $\overline{T}$; The formation of two Dirac points is a typical feature of WTIs \cite{FuPRB2007, FuPRB2007_2}. Although the Dirac points have not been directly detected, our data demonstrating the quasi-1D spin-momentum locking and its agreement with the first-principles calculations strongly support that the band S1 originates from the topological surface state of a WTI. 
The trivial bands (colored grey in Fig. 2k) crossing $E_\mathrm{F}$ in the calculations are from the dangling bonds at the \textit{a-b} surface, which will be removed in the calculations by proper atom absorption \cite{WangPRL2018} and do not appear in the surface Greens' function calculations \cite{WengPRX2014, ZhouSR2017} insensitive to the dangling bonds. They are not intrinsic, and not observed in the experiments. 

\begin{figure}[t]
\begin{center}
\includegraphics[width=0.45\textwidth]{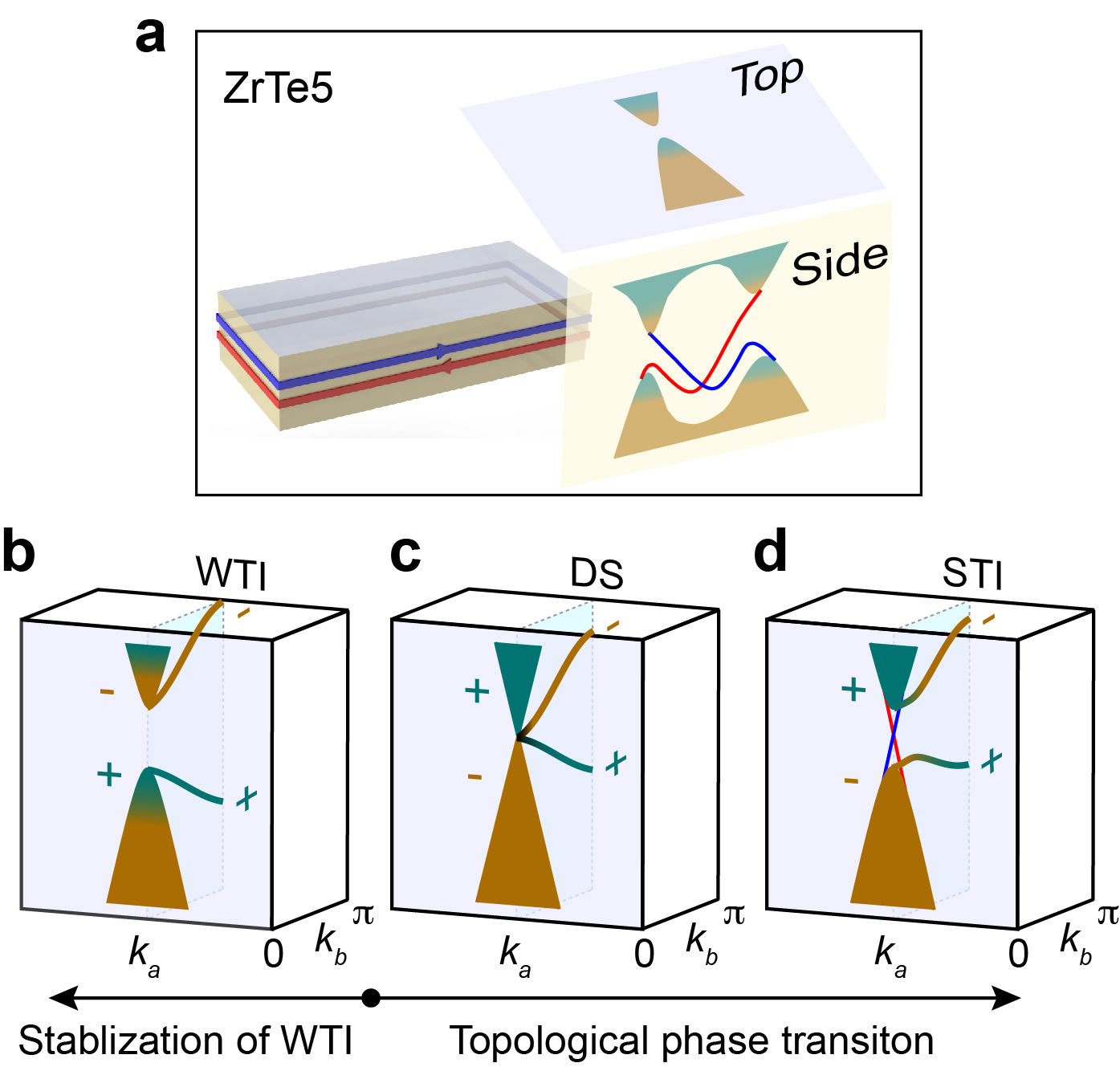}
\end{center}
 \caption{\label{cartoon}  \textbf{Weak topological insulator state of ZrTe$_5$ and control of its topological phase.} (a) In ZrTe$_5$, there are surface states on the side surface, while no surface state exists on the top surface. (b - d) Topological phase transition in ZrTe$_5$. The WTI state of ZrTe$_5$ can be stabilized by tensile strain (b), and a topological phase transition to DS (c) or even STI (d) can be obtained by compressive strain. There are two band inversions at k$_b$ = 0 and k$_b$ = $\pi$ in the WTI state of ZrTe$_5$. The band inversion at k$_b$ = 0 is switched by external strain in (b - d), while the band inversion at k$_b$ = $\pi$ has no change. }
\end{figure}

The small bulk gap ($\sim$ 32 meV) we observed by ARPES indicates that the WTI state in ZrTe$_5$ is protected only marginally from external perturbations. This is compatible with transport measurements showing semimetal behaviors in ZrTe$_5$ \cite{VallaNP2016, WangPRL2015, XiuNC2016, WangPNAS2017, OngNP2018}. Theoretical calculations proposed that the band gap changes with external strain \cite{WengPRX2014}. Such effect could be adopted to stabilize the topological state of mater in WTI, DS or even STI state; 
a relevant feature has been previously reported with the data showing a strain-dependent magneto-transport possibly caused by a band gap variation \cite{ChuSA2019}.  However, the signature of transport measurements is indirect, calling for the direct evidence of strain effect on the band structure by ARPES measurements\cite{ChiangNL2018}.  

In this study, the samples are glued on Ti or BeCu substrates, to which compressive or tensile strain is applied along the chain direction of ZrTe$_5$ (Fig. 3a; see Methods for more details). The band structures under different strain are displayed in Fig. 3b. The sample with no strain shows a band gap of $\sim$ 28 meV (Fig. 3b2), similar to that in Fig. 1. We have applied tensile strain to two samples, and find an increase of the band gap up to more than the double of the original value ($\sim61$ meV; Fig. 3b4), which makes a WTI state much more robust to external perturbations.  In contrast, we could instead reduce the gap value by the compressive strain,  even realizing a Dirac semimetal state with a gap no longer distinguishable (Fig. 3b1). We found that the cleaving process causes wrinkles in the sample under large compressive strain to relax the stress on the crystal lattice.  It thus becomes harder to apply larger compressive strain to realize a STI state by ARPES, which requires the sample cleavage. We note, however, that cleavage is not necessary for most applications, thus a STI state could be easily reached by a compression method. In Supplement Information Part VI and Fig. S6, we further show that the band gap can be reversibly controlled by external strain.

To better understand our data, we have carried out first-principles calculations of the band structure for different lattice constants of $a$ (Fig. 3c); In ARPES, the strain was applied to the samples along the chain direction (the $a$-axis). The calculations for compression by 0.25\% (Fig. 3c2) obtain a DS state, which is consistent with the ARPES results under compressive strain in Fig. 3b1.  On the other hand, the case of lattice expansion by 0.3\%  (Fig. 3c4) corresponds to our data under tensile strain in Fig. 3b3. 
Hence, we conclude that about 0.3\% compression and stretch are reached in our experiments with a strain device. This is consistent with the strain values measured by X-ray diffraction on the sample and strain gauge on the substrate (See Supplementary Information Part V and Fig. S5). 
The band gap variation with the lattice constant $a$ is summarized in Fig. 3d, where the experimental data roughly correspond to the solid dots.  The calculations further indicate that a STI state with a gap of 18 meV can be reached with 0.7\% compression (Fig. 3c1), and a WTI state with a band gap about 70 meV can be reached with 0.7\% stretch. 
Our experimental results provide the first direct evidence with ARPES on the band gap control and topological phase transition by external strain.

In summary, we have revealed that ZrTe$_5$ is a WTI, as sketched in Fig. 4a, by directly observing the electronic structure not only on the top surface but also on the side surface of crystal for the first time. 
On the top \textit{a-c} surface of ZrTe$_5$, the band structure was confirmed to be gapped at any temperatures lower than 200K.  In contrast, the side \textit{a-b} surface exhibits quasi-1D surface band with spin-momentum locked texture, which illustrates the WTI nature of ZrTe$_5$ distinct from a STI or a normal insulator. Under external strain, the band gap increases with tensile strain and decreases with compressive strain, providing a way to control the bulk gap of a WTI, or to realize a topological phase transition to a Dirac semimetal state, or even a STI state (Fig. 4b-d). A large band gap obtained by tensile strain can bring further protection of the 3D WTI state or monolayer QSH states in ZrTe$_5$ from thermal fluctuations or other external perturbations \cite{JHScience2018}. The quasi-1D surface states yielding the highly directional dissipationless spin current and the strain-tunable bulk gap make ZrTe$_5$ an ideal platform for 2D devices and spin engineering.

\textbf{Methods}

\textit{Sample growth}: 
High-quality crystals of ZrTe$_5$ were synthesized by the flux method with Te as the flux. High-purity elements (99.99999\% Te and 99.9999\% Zr) were loaded into a double-walled quartz ampoule and sealed under vacuum. The materials were first melted at 900 \textcelsius~in a box furnace and fully rocked for 72 h to achieve a homogeneous mixture. The melt was then slowly cooled and rapidly heated between 445 and 505 \textcelsius~for 21 days. Needle-like crystals were obtained. 

\textit{ARPES}: 
Laser-based ARPES and spin-resolved ARPES measurements were performed at the Institute for Solid State Physics, the University of Tokyo, with a laser delivering 6.994-eV photons. Photoelectrons were detected with ScientaOmicron R4000 analyzer and DA30L analyzer. The angle resolution was 0.3\textdegree~(0.7\textdegree) and the overall energy resolution was set to $\sim$ 5 meV (30 meV) in ARPES (spin-resolved ARPES) measurements. Three-dimensional spin-polarizations of photoelectrons were detected by two VLEED-type spin detectors\cite{YajiRSI2016}. Synchrotron-based nano-ARPES measurements were performed at the 3.2L-Spectromicroscopy beamline of the Elettra Light Source. A Schwarzschild objective was used to focus the photon beam to a spot $<$ 1$\mu$m in size. The photon energy was set to 74 eV and temperature was set to around 100 K. The overall energy resolution was set to be better than 60 meV.

\textit{Band fitting:} 
The band dispersions are extracted from MDC peaks. To avoid any complication, we simply used the position of the MDC maximums. Since the band dispersion highly resembles a cone structure, we used the following formula to fit the band structure:
\begin{equation*}
f(x)=\left\{\begin{matrix}
c_1 (x - c4) + c_3, & x \ge c_4 \\ 
c_2 (x - c4) + c_3, & x<c_4
\end{matrix}\right.
\end{equation*}
where $c_1, c_2, c_3, c_4$ are the fitting parameters. If the computer program returns 0 or 1 for the inequality $x \ge c_4$ and $x<c_4$, the above formula can be rewritten in one line:
\begin{equation*}
f(x)= [c_1 (x - c4) + c_3] (x \ge c_4) + [c_2 (x - c4) + c_3] (x<c_4)
\end{equation*}
Since the band dispersion may not be perfectly symmetric, two independent slopes were used for a better fitting. We note this method will underestimate the gap size, as the band top is actually curved due to the band gap opening. However, since the band gap is very small and the band dispersion resembles a cone structure to the very top, the difference should be quite small. Also we mainly compare the gap change with strain, such underestimation should not affect the comparison. 

\textit{Strain device:} 
The strain device shown in Fig. 3a has a unibody structure made of Be-Cu. There are two walls on the flat plate: a thick one (left) and a thin one (right). A substrate is fixed on the top of the two walls. By reducing or increasing the space between the two walls, we can apply compressive or tensile strain on the substrate, thus on the sample. In the device to apply compressive strain, the thick wall (left) has a through hole, while the thin wall (right) has a threaded hole. Thus by tightening the screw, the space between the two walls will be reduced, and compressive strain will be applied. In the device to apply tensile strain, the thick wall has a threaded hole, while the thin wall is blinded. Thus by the tightening the screw, the space will be enlarged, and tensile strain will be applied. Due to the limited space of the sample holder, Torr Seal is used to fix the substrate on top of the walls. 

The data in Fig. 3b are all taken from one thick sample. We exfoliated this thick sample with scotch tape and got different flakes. Then we mounted these different flakes to the substrates of the strain devices, and carried out the experiments. With this method, we avoided the problem that different samples may have slightly different bulk band gap. 

\textit{DFT calculations:} 
Our calculations have been performed with the projector augmented wave method implemented in Vienna ab initio simulation package \cite{Kresse1996, KressePRB1996}. Generalized gradient approximation of Perdew-Burke-Ernzerhof type is used \cite{PerdewPRL1996}. The k-point sampling grids are set to 13$\times$13$\times$7 and 7$\times$7$\times$1 for ZrTe$_5$ bulk and  ZrTe$_5$ slab, respectively. Spin orbital coupling (SOC) is included in the calculations. Considering the fact that ZrTe$_5$ is a strong topological insulator \cite{WengPRX2014} when the experimental lattice constants \cite{FjellvagSSC1986} are used, we take a slightly increased interlayer distance (0.25 \AA) to mimic the weak topological insulator phase, which is consistent with our experimental observations. The ionic positions are relaxed until force on each ion is less than 0.01 eV/\AA.

\textbf{Data availability}
The data that support the findings of this study are available from the corresponding authors upon reasonable request.

\begin{addendum}
\item We acknowledge X.X. Wu and Y.M. Li for useful discussions. P. Zhang is an International Research Fellow of the Japan Society for the Promotion of Science. This work was supported by MEXT Quantum Leap Flagship Program (MEXT Q-LEAP, Grant No. JPMXS0118068681) and by the JSPS KAKENHI (Grants No. JP18H01165, JP19H02683, JP19F19030, and JP19H00651). Work at Brookhaven National Laboratory was supported  by the U.S. Department of Energy, Office of Science,  Office of Basic Energy Sciences, under Contract No. DOE SC0012704.
\item[Competing Interests] The authors declare that they have no competing financial interests.
\item[Correspondence] Correspondence and request for materials should be addressed to P.Z. or T.K. (emails: zhangpeng@issp.u-tokyo.ac.jp, kondo1215@issp.u-tokyo.ac.jp)
\item[Author contributions] P.Z. did the ARPES measurements and analyzed the data with help from R.N., K.Kuroda., C.L., K.Kawaguchi, K.Y., A.H., M.L., V.K., A.G., A.B. and S.S.. S.N. and H.W. did the theory calculations. Q.L. and G.D.G. synthesized the samples. All authors discussed the paper. P.Z. and T.K. wrote the manuscript and supervised the project.
\end{addendum}

\end{document}